\documentclass[aps,prl,twocolumn,longbibliography]{revtex4}
\usepackage{amsfonts,amssymb,amscd,amsthm}
\usepackage{graphicx}
\usepackage{mathrsfs}
\usepackage[intlimits]{amsmath}
\usepackage[colorlinks, citecolor=red]{hyperref}
\usepackage{physics}
\usepackage{units}
\newcommand{\Yb}{\ensuremath{^{171}\mathrm{Yb}^+~}}
\newcommand{\nm}{\unit{nm}}
\newcommand{\MHz}{\unit{MHz}}
\newcommand{\THz}{\unit{THz}}
\newcommand{\GHz}{\unit{GHz}}
\newcommand{\kHz}{\unit{kHz}}
\newcommand{\Hz}{\unit{Hz}}
\newcommand{\ms}{\unit{ms}}
\newcommand{\us}{\unit{\mu s}}
\begin{document}

\title{Observation of a quantum phase transition in the quantum Rabi model with a single trapped ion}
\author{M.-L. Cai\footnote{These authors contribute equally to this work}$^{1}$, Z.-D. Liu$^{*1}$, W.-D. Zhao$^{*1}$, Y.-K. Wu$^{1}$, Q.-X. Mei$^{1}$, Y. Jiang$^{1}$, L. He$^{1}$, X. Zhang$^{2,1}$, Z.-C. Zhou$^{1,3}$, L.-M. Duan\footnote{Corresponding author: lmduan@tsinghua.edu.cn}$^{1}$}
\affiliation{$^{1}$Center for Quantum Information, Institute for Interdisciplinary Information Sciences, Tsinghua University, Beijing 100084, PR China}
\affiliation{$^{2}$Department of Physics, Renmin University, Beijing 100084, PR China}
\affiliation{$^{3}$Beijing Academy of Quantum Information Sciences, Beijing 100193, PR China}

\begin{abstract}
\section{Abstract}
Quantum phase transitions (QPTs) are usually associated with many-body systems in the thermodynamic limit when their ground states show abrupt changes at zero temperature with variation of a parameter in the Hamiltonian. Recently it has been realized that a QPT can also occur in a system composed of only a two-level atom and a single-mode bosonic field, described by the quantum Rabi model (QRM). Here we report an experimental demonstration of a QPT in the QRM using a trapped ion. We measure the spin-up state population and the average phonon number of the ion as two order parameters and observe clear evidence of the phase transition via adiabatic tuning of the coupling between the ion and its spatial motion. An experimental probe of the phase transition in a fundamental quantum optics model without imposing the thermodynamic limit opens up a window for controlled study of QPTs and quantum critical phenomena.
\end{abstract}

\maketitle

\section{Introduction}
Quantum phase transitions (QPTs) have become one of the focuses of condensed matter physics. Unlike classical phase transitions that occur at finite temperature, a QPT can occur at zero temperature under quantum fluctuations \cite{RevModPhys.69.315,Vojta_2003,sachdev2011quantum}. When a control parameter, such as the external magnetic field or the doping of a component, is scanned across a quantum critical point, the ground state of the system changes abruptly, characterized by a spontaneous symmetry breaking or a change in the topological order \cite{Vojta_2003,RevModPhys.75.913}.

Studies of QPTs usually consider many-body systems in the thermodynamic limit, with the particle number $N$ approaching infinity \cite{sachdev2011quantum}. However, it was recently realized that a QPT can also occur in a small system with only two constituents, a two-level atom and a bosonic mode, described by the quantum Rabi model (QRM) \cite{PhysRevA.85.043821,PhysRevB.69.113203,PhysRevA.70.022303,PhysRevA.81.042311,PhysRevA.82.025802,PhysRevA.87.013826,PhysRevLett.115.180404,PhysRevLett.118.073001} which is one of the simplest models of light-matter interactions. Its Hamiltonian can be expressed as (throughout this paper we set $\hbar=1$ for simplicity)
\begin{equation}
\label{eq:rabi_model}
\hat{H}_{\mathrm{QRM}}=\frac{\omega_\mathrm{a}}{2} \hat{\sigma}_{z}+\omega_\mathrm{f} \hat{a}^{\dagger} \hat{a}+\lambda\left(\hat{\sigma}_{+}+\hat{\sigma}_{-}\right)\left(\hat{a}+\hat{a}^{\dagger}\right),
\end{equation}
where $\hat{a}^{\dagger}$ ($\hat{a}$) is the bosonic mode creation (annihilation) operator and $\hat{\sigma}_+$ ($\hat{\sigma}_-$) is the two-level system raising (lowering) operator; $\omega_\mathrm{a}$, $\omega_\mathrm{f}$ and $\lambda$ are the atomic transtion frequency, the field mode frequency and the coupling strength between the two subsystems, respectively. This model has been widely studied in multiple paramter regions with many experimental platforms. When $|\omega_\mathrm{a}-\omega_\mathrm{f}|\ll|\omega_\mathrm{a}+\omega_\mathrm{f}|$ and $\lambda/\omega_\mathrm{f}\ll1$ are fulfilled, the rotating wave approximation (RWA) can be used to simplify the QRM to the Jaynes-Cummings model (JCM) \cite{1443594,Pedernales2015} which has been investigated first in cavity QED \cite{Miller_2005,Walther_2006,RevModPhys.73.565} and trapped ions \cite{RevModPhys.75.281}, and then in other platforms such as quantum dots \cite{RevModPhys.79.1217} and circuit QED \cite{Wallraff2004,Devoret1169}. When $\lambda$ becomes comparable to $\omega_\mathrm{a}+\omega_\mathrm{f}$, the RWA breaks down leading to the ultra-strong coupling regime ($\lambda/\omega_\mathrm{f} \gtrsim 0.1$) and deep-strong coupling regime ($\lambda/\omega_\mathrm{f} \gtrsim 1$) \cite{Pedernales2015}. Many exotic dynamical properties in these regimes have been observed recently in a plenty of quantum systems such as circuit QED \cite{PhysRevLett.105.237001,Niemczyk2010,Braumuller2017,Forn-Diaz2017,Yoshihara2017,Langford2017}, photonic system \cite{PhysRevLett.108.163601}, semiconductor system \cite{PhysRevLett.102.186402,Gunter2009} and trapped ions \cite{PhysRevX.8.021027}.

In the trapped-ion systems, previous works on the simulation of the QRM have been performed in various regimes. For $\omega_\mathrm{a}=0,\omega_\mathrm{f}\neq0$, the QRM reduces to the spin-dependent force Hamiltonian which is crucial in trapped-ion quantum computation \cite{SDF,Sackett2000,PhysRevA.72.062316,PhysRevLett.112.190502}. For $\omega_\mathrm{a}\neq0,\omega_\mathrm{f}=0$, the Dirac equation has been simulated with trapped ions \cite{Gerritsma2010,PhysRevLett.106.060503}. For $\omega_\mathrm{a}=0,\omega_\mathrm{f}=0$, the coupling-only regime can be realized and it has been exploited to engineer the Schr\"odinger cat state \cite{Lo2015,PhysRevLett.116.140402} and the grid state \cite{PhysRevX.8.021001,Fluhmann2019}. By controlling the experimental parameters, Ref.~\cite{PhysRevX.8.021027} has access to the ultra-strong and the deep-strong coupling regimes. However, most of the previous works focus on the evolution dynamics governed by the QRM Hamiltonian in multiple regimes.

Our work realizes the model Hamiltonian in a special parameter region $\omega_\mathrm{a}\gg\omega_\mathrm{f}$, which allows the study of a QPT with the phases controlled by the coupling strength $\lambda$ in the QRM. In Ref.~\cite{PhysRevLett.115.180404}, it has been shown that an order parameter, the rescaled photon number in the bosonic mode, is shown to stay zero in the normal phase while acquiring positive values in the superradiant phase with a spontaneous breaking of the $Z_2$ parity symmetry. The ground state of the system exhibits nonanalytical behavior at the critical point, supporting a second-order phase transition at zero temperature \cite{PhysRevLett.115.180404}. We experimentally demonstrate this type of QPT without the conventional thermodynamic limit of a large number of particles. Through laser driving near the blue and the red motional sidebands, we use a single trapped $\Yb$ ion to simulate the QRM Hamiltonian with adjustable parameters \cite{Pedernales2015,PhysRevX.8.021027}. We perform a slow quench on the control parameter and measure the average atomic-level population and the average phonon number as the order parameters on both sides of the transition point. The experiments are repeated for the increasing ratios of $\omega_\mathrm{a}$ and $\omega_\mathrm{f}$, with the limit $\omega_\mathrm{a}/\omega_\mathrm{f}\to\infty$ analogous to the thermodynamic limit \cite{PhysRevLett.115.180404}. From the qualitative behavior of the order parameters under the increasing ratios, we obtain strong evidence of the QPT in the QRM, although the ratio parameter is still not large enough for a precise scaling analysis of the critical phenomenon.
Our work simulates the QRM in a special parameter region and develops a tool for adiabatic passages that allows the controlled study of a QPT, and showcases the possibility of exploring the universal QPT properties using the trapped-ion system, which has a number of tunable experimental knobs that can be used for a controlled study of the QPT and the critical phenomena under influence of various effects.

\section{Results}

\textbf{The quantum critical point in the quantum Rabi model.}
To study the QPT, the low-energy effective Hamiltonian in the limit $\omega_\mathrm{a}/\omega_\mathrm{f}\to\infty$ has been derived in Ref.~\cite{PhysRevLett.115.180404}. When the control parameter $g\equiv2\lambda/\sqrt{\omega_\mathrm{a}\omega_\mathrm{f}} < 1$, the effective Hamiltonian in the normal phase is given by $\hat{H}_{\mathrm{np}}=\omega_\mathrm{f} \hat{a}^{\dagger} \hat{a}- g^{2}\omega_\mathrm{f} (\hat{a}+\hat{a}^{\dagger})^{2} / 4 - \omega_\mathrm{a}/2$ with the qubit frozen in its ground state;
and when $g>1$ we have the effective Hamiltonian in the superadiant phase $\hat{H}_{\mathrm{sp}} = \omega_\mathrm{f} \hat{a}^{\dagger} \hat{a} - \omega_\mathrm{f}(\hat{a}+\hat{a}^{\dagger})^{2}/(4 g^{4})- \omega_\mathrm{a} (g^{2}+g^{-2})/4$ in a displaced frame of the bosonic mode, with the qubit ground state now rotated toward the x axis due to its strong coupling to the bosonic mode. This generates non-zero spin and bosonic population in the ground state of the superradiant phase. Hence, we can utilize both the rescaled bosonic mode number ($n_{\mathrm{f}}\equiv(\omega_\mathrm{f} / \omega_\mathrm{a}) \langle \hat{a}^{\dagger} \hat{a} \rangle$) and the spin population ($n_\mathrm{a}=1+\langle \hat{\sigma}_{z} \rangle$) at ground state as the order parameters: in the limit $\omega_\mathrm{a}/\omega_\mathrm{f}\to \infty$, we have $n_\mathrm{f}=0 (n_\mathrm{a}=0)$ when $g<1$ and $n_\mathrm{f}=(g^{4}-g_{\mathrm{c}}^{4}) / (4 g^{2}) (n_{\mathrm{a}}=1-g^{-2})$ for $g>1$ \cite{PhysRevLett.115.180404,PhysRevLett.118.073001}.

\begin{figure*}[tbp]
   \includegraphics[width=0.95\linewidth]{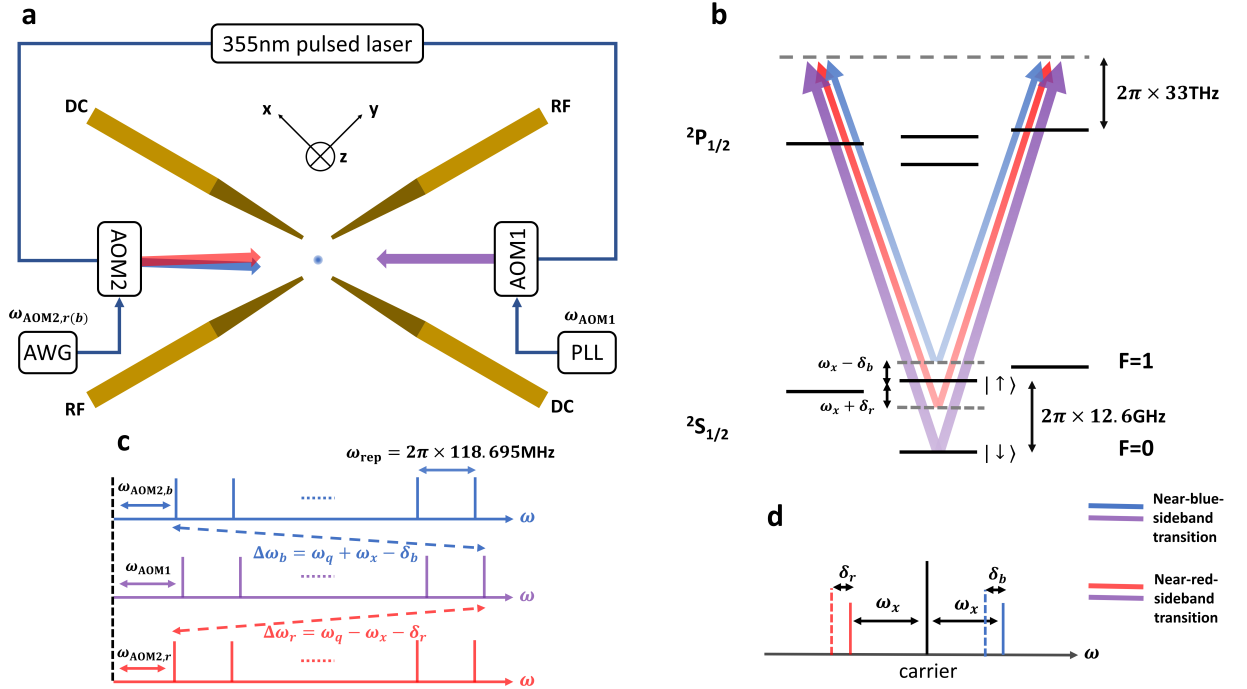}
   \caption{\textbf{Schematic for experimental observation of QPT in the quantum Rabi model.} \textbf{a}. Schematic experimental setup. The $\Yb$ ion is confined in the middle of a four-blade Paul trap, with the principal axes of the secular motion along the $x$, $y$ and $z$ directions. Two counter-propagating $355\,\nm$ pulsed-laser beams are focused on the ion, with a nonzero differential wave vector component along the $x$ direction. The two laser beams are controlled by two acousto-optic modulators (AOMs). AOM1 is driven by a radio-frequency (RF) signal from a phase-locked loop (PLL) \cite{PLL} and AOM2 is controlled by an arbitrary waveform generator (AWG). \textbf{b}. Schematic level structure of $\Yb$. The two qubit states are two ${}^2S_{1/2}$ hyperfine ground states $\ket{\uparrow}=\ket{F=1,m_F=0}$ and $\ket{\downarrow}=\ket{F=0,m_F=0}$, at the separation $\omega_q\approx 2\pi\times 12.6\,\GHz$. The Raman transition is mediated by a virtual level about $2\pi \times 33\,\THz$ above the ${}^2P_{1/2}$ levels.
   The differential frequencies of the laser beams are tuned close to the blue and the red motional sidebands, i.e. $\omega_{x}-\delta_b$ and $-(\omega_{x}+\delta_r)$ from the carrier transition. The legend at lower right shows clearly that the purple beam and the blue (red) beam form a near-blue-sideband (near-red-sideband) Raman transition.
   \textbf{c}. The $355\,\nm$ pulsed laser has a frequency-comb structure \cite{frequencycomb} with the repetition rate $\omega_{\mathrm{rep}}\approx 2\pi\times 118.695\,\MHz$. With small frequency adjustments in the AOMs, the desired Raman transitions can be achieved between distant teeth of the frequency combs. \textbf{d}. Relative positions of the carrier transition (black) and two motional sidebands (red and blue) in solid lines and the bichromatic Raman-transition frequencies (red and blue) in dashed lines.}
   \label{Fig 0}
\end{figure*}

\textbf{Experimental setup.}
We use a single $\Yb$ ion confined in a linear Paul trap to simulate the QRM, as shown in Fig.~\ref{Fig 0}a.
By performing the Doppler cooling followed by a resolved sideband cooling \cite{RevModPhys.75.281}, the spatial motion of the ion along one of its principal axes $x$, with the frequency $\omega_x=2\pi\times 2.35\,\MHz$, is cooled close to the ground state. Its motional degree of freedom can be well described as a quantum harmonic oscillator, and thus serves as the bosonic mode in the QRM.
The two hyperfine states in the ground-state manifold ${}^2S_{1/2}$ are chosen as the qubit states, i.e. $\ket{\uparrow}=\ket{F=1,m_F=0}$ and $\ket{\downarrow}=\ket{F=0,m_F=0}$, with a frequency difference $\omega_q\approx 2\pi\times 12.6\,\GHz$ as shown in Fig.~\ref{Fig 0}b.
We use two counter-propagating $355\,\nm$ pulsed-laser beams to manipulate the hyperfine qubit through Raman transition. The pulsed laser has a frequency-comb structure as shown in Fig.~\ref{Fig 0}c, which can help bridge the large frequency gap $\omega_q$ between the two
levels \cite{frequencycomb}; the undesired teeth of the frequency combs can effectively produce a fourth-order AC Stark shift \cite{PhysRevA.94.042308}, which we carefully measure and compensate in the experiment (see Methods for more details). Two acousto-optic modulators (AOMs) are used to fine-tune the frequencies and the amplitudes of the laser beams for driving the Raman transition.

The orientation of the laser beams are chosen such that there is a nonzero differential wave vector component $\Delta k_x$ along the $x$ axis. Let us first consider a single pair of Raman beams with the frequency and the phase difference $\Delta\omega$ and $\Delta\phi$ generating a Rabi frequency $\Omega$. The laser-ion coupling Hamiltonian is given by $\hat{H}_{\mathrm{couple}}=\Omega \cos (\Delta k_x \cdot \hat{x}-\Delta \omega \cdot t+\Delta\phi) \hat{\sigma}_{x}$ \cite{transverse-phonon-modes}, where $\hat{x}=x_0(\hat{a}+\hat{a}^{\dagger})$ is the ion-position operator with $x_0$ being the ground state wave-packet width. Considering the Lamb-Dicke approximation $\eta\sqrt{2\bar{n}+1}\ll 1$ where $\eta\equiv \Delta k_x x_0$ is the Lamb-Dicke parameter and $\bar{n}$ is the average phonon number of the motional state (see Supplementary Information for more details about the correction of the Lamb-Dicke approximation), we transfer $\hat{H}_{\mathrm{couple}}$ into the interaction picture of the uncoupled Hamiltonian $\hat{H}_{0}=\omega_{q}\hat{\sigma}_{z} / 2 +\omega_{x} \hat{a}^{\dagger} \hat{a}$, and get the interaction Hamiltonian
$\hat{H}_r=(\eta \Omega_r/2) (\hat{a} \hat{\sigma}_{+} e^{i \delta_{r} t}+\hat{a}^{\dagger} \hat{\sigma}_{-} e^{-i \delta_{r} t})$ if the frequency difference $\Delta\omega$ is tuned close to the red motional sideband with $\delta_{r}=\omega_q-\omega_x-\Delta\omega$,
and $\hat{H}_b=(\eta \Omega_b/2) (\hat{a}^{\dagger} \hat{\sigma}_{+} e^{i \delta_{b} t} + \hat{a} \hat{\sigma}_{-} e^{-i\delta_{b} t})$ when $\Delta\omega$ is tuned close to the blue sideband with $\delta_{b}=\omega_q+\omega_x-\Delta\omega$.

In order to construct the QRM Hamiltonian, we employ the bichromatic Raman beams as shown in Fig.~\ref{Fig 0} driving the red and the blue sidebands simultaneously \cite{Sackett2000,SDF,PhysRevA.72.062316} using the specific implementation proposed and realized recently in Ref.~\cite{Pedernales2015,PhysRevX.8.021027}, as shown in Fig.~\ref{Fig 0}b. If we set the two Rabi frequencies to be the same $\Omega_r=\Omega_b=\Omega$ (in the experiment we can calibrate them such that the imbalance  $\abs{\Omega_{r}-\Omega_{b}}/\abs{\Omega_r+\Omega_{b}} \le 2\%$),
the resulting Hamiltonian is $\hat{H}_{rb}=(\eta\Omega/2)\hat{\sigma}_{+} (\hat{a} e^{i \delta_r t} + \hat{a}^{\dagger} e^{i\delta_b t})+h.c.$ which corresponds to the interaction picture Hamiltonian with respect to the uncoupled Hamiltonian $\hat{H}_0'=-(\delta_{b}+\delta_{r})\hat{\sigma}_{z} / 4 - (\delta_{b}-\delta_{r})\hat{a}^{\dagger} \hat{a} / 2$ \cite{Pedernales2015},
\begin{equation}
\begin{aligned}
\label{6}
\hat{H}^I_{rb}=& \frac{\delta_{b}+\delta_{r}}{4} \hat{\sigma}_{z}+\frac{\delta_{b}-\delta_{r}}{2} \hat{a}^{\dagger} \hat{a}\\
& +\frac{\eta \Omega}{2}\left(\hat{\sigma}_{+}+\hat{\sigma}_{-}\right)\left(\hat{a}+\hat{a}^{\dagger}\right).
\end{aligned}
\end{equation}
We clearly see the transformed Hamiltonian is exactly the QRM Hamiltonian if we identify $\omega_{\mathrm{a}}=(\delta_{b}+\delta_{r})/2$, $\omega_{\mathrm{f}}=(\delta_{b}-\delta_{r})/2$ and $\lambda=\eta\Omega/2$. From our definition, the control parameter is $g\equiv 2\lambda/\sqrt{\omega_\mathrm{a}\omega_\mathrm{f}}=2\eta\Omega/\sqrt{\delta_b^2-\delta_r^2}$. Since the uncoupled Hamiltonian $\hat{H}_0'$ commutes with our desired observables, the spin ($\hat{\sigma}_z$) and the phonon ($\hat{a}^{\dag}\hat{a}$) population, their measurements will not be affected by this transformation \cite{Pedernales2015}. By controlling the experimental parameters $\delta_b$, $\delta_r$ and $\Omega$, we can achieve the simulation in the regime $\omega_\mathrm{a}\gg\omega_\mathrm{f}$ where an observation of a QPT is possible.

\begin{figure}[tbp]
   \includegraphics[width=0.95\linewidth]{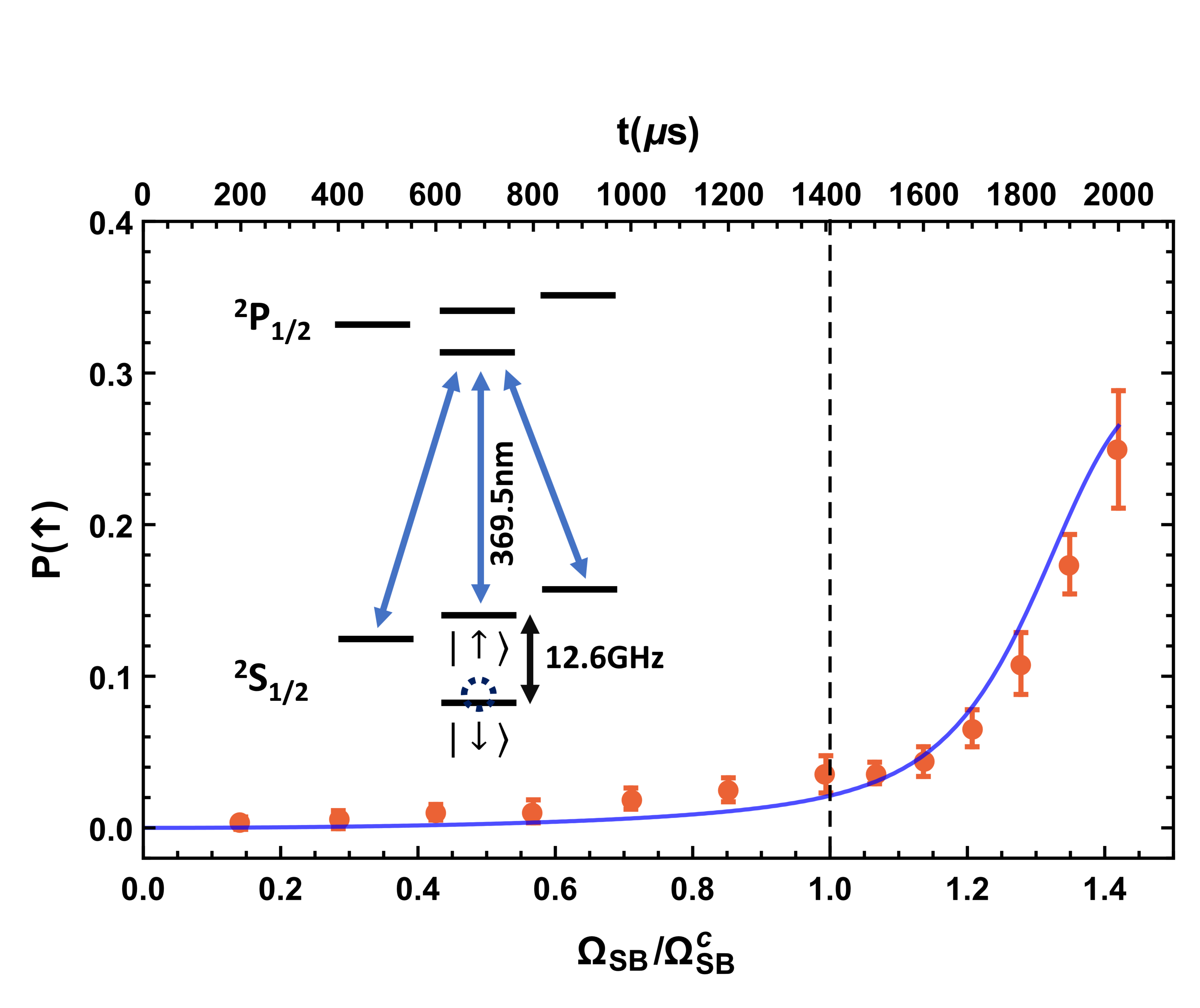}
   \caption{\textbf{Spin-up state population versus sideband Rabi frequency.} By setting $\delta_b=2 \pi\times 52.0\,\kHz$ and $\delta_r=2\pi\times48.0\,\kHz$, we keep the ratio parameter $R=\omega_\mathrm{a}/\omega_\mathrm{f}=25$ fixed. As we increase the sideband Rabi frequency $\Omega_{\mathrm{SB}}$ (bottom axis) linearly with time (top axis), i.e $\Omega_{\mathrm{SB}}=\Omega_{\mathrm{max}}t/\tau_{q}$ where $\Omega_{\mathrm{max}}=2\pi\times 14.2\,\kHz$ and the quench time $\tau_q=2\,\ms$ are two pre-determined parameters, the control parameter $g(t)=2\Omega_{\mathrm{SB}}(t)/\sqrt{\delta_b^2-\delta_r^2}$ goes up accordingly. With a duration time $t$, we prepare a target state under $g(t)$ and measure the spin-up state population by florescence detection.
   Every orange dot is the average of 20 rounds of measurements of the spin-up state population, corrected by subtracting the $1.0\%$ dark-state detection error as the background; the error bar is estimated as one standard deviation of the 20-round outcomes (see Supplementary Information for more details about the error bar estimation). For each round of measurement, we repeat the experiment sequence for 500 shots and take the average. The blue curve is the theoretical value by directly solving the time-dependent Schr\"odinger equation under the QRM Hamiltonian. The vertical dashed line is an indication of the quantum critical point $g_\mathrm{c}=1$ (corresponding to $\Omega_{\mathrm{SB}}^\mathrm{c}=2\pi\times 10\,\kHz$). The inset shows the florescence detection scheme of $\Yb$ ions \cite{PhysRevA.76.052314}.}
   \label{Fig 1}
\end{figure}

\textbf{Observation of quantum phase transition from the spin population.}
To observe the QPT from the normal phase to the superradiant phase in the QRM Hamiltonian,
we consider two measurable order parameters, the spin-up state population $(1+\langle \hat{\sigma}_{z} \rangle)/2$ \cite{PhysRevLett.118.073001} and the average phonon number $\langle\hat{a}^\dag \hat{a}\rangle$ \cite{PhysRevLett.115.180404}. As the control parameter $g$ rises from zero to above the quantum critical point,
the $Z_2$ parity symmetry is broken and these two values at the ground state will accordingly increase from zero to a non-zero value.
However, it is hard to prepare the ground state of a general Hamiltonian \cite{doi:10.1137/080734479}, and since the energy gap closes at the quantum critical point, we are not able to adiabatically scan the control parameter across this point without generating the quasi-particle excitations into the system \cite{PhysRevLett.115.180404}. Therefore, in this experiment we perform slow quench on the control parameter as suggested by Ref.~\cite{PhysRevLett.115.180404}, and compare the measured values with the theoretical predictions.

First we set $\delta_b=2 \pi\times 52.0\,\kHz$ and $\delta_r=2\pi\times48.0\,\kHz$, which corresponds to a ratio $R\equiv \omega_{\mathrm{a}}/\omega_{\mathrm{f}}=25$ between the atomic transition frequency and the field mode frequency in the QRM. Under this finite ratio, the energy gap at the quantum critical point becomes finite which is around $0.4\omega_{\mathrm{f}}=2\pi\times0.8\,\kHz$ \cite{PhysRevLett.115.180404}, indicating that the quench time should at least be $1.25\,\ms$ such that the prepared state does not deviate too much from the true ground state.
After sideband cooling, we initialize the ion in the ground state $\ket{\downarrow,n=0}$. Then we linearly increase the sideband Rabi frequency such that $\Omega_{\mathrm{SB}}(t)\equiv\eta\Omega(t)=\Omega_{\mathrm{max}} t/\tau_{q}$ where $\Omega_{\mathrm{max}}=2\pi\times 14.2\,\kHz$ and the quench time $\tau_q=2\,\ms$ are two pre-determined parameters. In other words, the time to reach the critical point $\Omega_{\mathrm{SB}}^\mathrm{c}=\sqrt{\delta_b^2-\delta_r^2}/2=2\pi\times 10\,\kHz$ is about $1.4\,\ms$. We expect the quantum state of the system to follow the slow quench of the control parameter $g(t)=\Omega_{\mathrm{SB}}(t) / \Omega_{\mathrm{SB}}^\mathrm{c}$. Hence with a duration time $t$, we generate the target state under a specific coupling strength of the QRM and measure the order parameters.

The spin-up state population can be measured by a resonant driving on the $|{}^2S_{1/2},F=1\rangle \to |{}^2P_{1/2},F=0\rangle$ cyclic transition of the $\Yb$ ion and a detection of the scattered photon counts \cite{PhysRevA.76.052314}. The result is shown in Fig.~\ref{Fig 1}.
Every orange data point is the average of 20 rounds of measurements of the spin-up state population and has been corrected by subtracting the $1.0\%$ dark-state detection error which arises from the small residual off-resonant coupling of the detection laser to the bright state \cite{PhysRevA.76.052314} as the background. For each round of measurement, the outcome is acquired by averaging over 500 shots of the experiment sequence. The error bar is estimated by one standard deviation of the 20 rounds. We clearly observe the increase of the order parameter $(1+\langle \hat{\sigma}_{z}\rangle)/2$ after the quantum critical point (the vertical dashed line in Fig.~\ref{Fig 1}) despite the relatively low sharpness due to the finite ratio parameter $R$, which agrees well with the numerical simulation (the blue curve in Fig.~\ref{Fig 1} from numerically solving the time-dependent Schr\"odinger equation of the QRM Hamiltonian).

\begin{figure*}[tbp]
   \includegraphics[width = 0.95\linewidth]{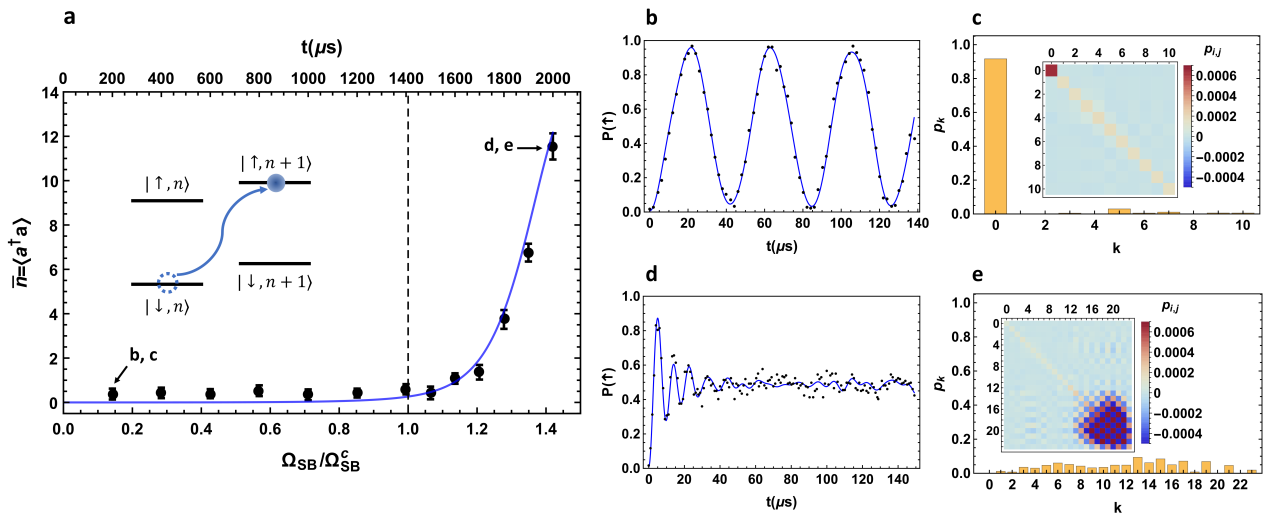}
   \caption{\textbf{Average phonon number versus sideband Rabi frequency.} Again we set $\delta_b=2 \pi\times 52.0\,\kHz$ and $\delta_r=2\pi\times48.0\,\kHz$, thus the ratio parameter $R=\omega_\mathrm{a}/\omega_\mathrm{f}=25$. With the same quench process as above, we prepare the target states and measure the corresponding average phonon numbers. \textbf{a}. Each black dot is a measured average phonon number for a specific ground state. Its value and the error bar are determined according to (b-e). The blue curve is the theoretical result by solving the time-dependent Schr\"odinger equation. The inset shows the blue sideband scheme for analyzing the phonon number distribution: before the measurement, we optically pump the spin state into $\ket{\downarrow}$ \cite{PhysRevA.76.052314} with tiny influence to the phonon state population; then we drive the blue sideband transition for various time interval and fit the obtained spin-up state population to extract the phonon distribution. For the leftmost data point in the normal phase, \textbf{b} presents the experimental data (black dots, averaged over 200 shots) and the fitted curve (blue line), and \textbf{c} shows the fitted population $p_k$ ($k=0,\,1,\,\cdots$) with the covariance matrix shown in the inset. The error bar in (a) is computed from this covariance matrix as one standard deviation for the average phonon number. Similarly \textbf{d} and \textbf{e} show the results for the rightmost data point in (a) in the superradiant phase. More details can be found in Methods.}
   \label{Fig 2}
\end{figure*}

\textbf{Observation of quantum phase transition from the phonon number.}
Next we consider another order parameter, the average phonon number. After the slow quench of the QRM Hamiltonian, a short optical pumping pulse of $5\,\us$ is applied to pump the internal state of the ion (qubit state) into $\ket{\downarrow}$ \cite{PhysRevA.76.052314} with negligible effect on the motional state (phonon state) population. Then we drive the blue-sideband transition between $\ket{\downarrow,n}$ and $\ket{\uparrow,n+1}$ ($n=0,\,1,\,\cdots$) for various time interval $t$. By fitting the resultant spin-up state population, we can reconstruct the population of different phonon states, thus calculating the average phonon number \cite{PhysRevLett.76.1800,RevModPhys.73.565,Hofheinz2008,Hofheinz2009,PhysRevLett.76.1796,RevModPhys.75.281,Kienzler53,Lo2015,PhysRevLett.116.140402}.

With the same experimental parameters as above, the results are shown in Fig.~\ref{Fig 2}. Each black dot in Fig.~\ref{Fig 2}a is the calculated average phonon number from the phonon population distribution with the error bar estimated by one standard deviation. In Fig.~\ref{Fig 2}b we show an example for the blue sideband signal of the leftmost data point in Fig.~\ref{Fig 2}a. The measured spin-up state population is fitted by the blue curve to give the phonon state population $\{p_k\}$ ($k=0,\,1,\,\cdots$) with a suitable truncation. The fitting result is shown in Fig.~\ref{Fig 2}c with a covariance matrix (inset) representing the correlation between different $p_k$'s, from which we further deduce the standard deviation of the average phonon number, assuming a joint Gaussian distribution \cite{Nonlinear}. More details can be found in Methods. As we can see, for this data point we get a very low average phonon number, consistent with the fact that it is deep in the normal phase. Similarly, Fig.~\ref{Fig 2}d and Fig.~\ref{Fig 2}e show the results for the rightmost data point in Fig.~\ref{Fig 2}a. Here we get much faster oscillation at the beginning of the blue sideband data owing to the much higher phonon number population (the sideband Rabi oscillation frequency $\sim \sqrt{n+1}\eta\Omega$) in the superradiant phase, as well as much faster decay since the phonon number has a wider distribution. In this case we get larger uncertainty in each fitted $p_k$. However, they are strongly correlated as shown by the off-diagonal elements of the covariance matrix (inset of Fig.~\ref{Fig 2}e), and we still get a reasonable error bar for the average phonon number. Finally, in Fig.~\ref{Fig 2}a we further compare the measured average phonon number with the theoretical values from numerically solving the time-dependent Schr\"odinger equation. Again these results agree well within the error bars.

It should be pointed out that the fourth order AC Stark shift induced by the laser beams is not zero in our setup \cite{PhysRevA.94.042308}, and will increase as we gradually turn up the coupling strength of the QRM in the above experiments. Therefore they cannot be compensated by a static frequency shift in the laser beams, but require a dynamic compensation by phase modulation of the laser as shown in Methods.
Also note that for our slow quench dynamics to maintain quantum coherence, the total quench time $\tau_q$ should be shorter than the motional decoherence time $\tau_d$ of the trapped ion.
The motional coherence of our system is largely affected by the $50\,\Hz$ noise from the AC power line. Therefore we use a line-trigger to lock the experimental sequence to the AC signal from the power line, which extends the motional decoherence time to over $5\,\ms$.

\begin{figure*}[tbp]
   \includegraphics[width = 0.95\linewidth]{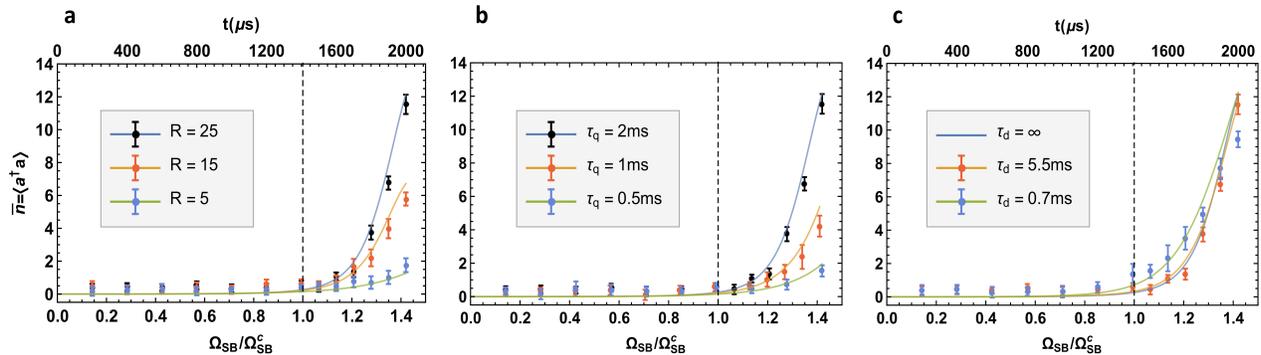}
   \caption{\textbf{Average phonon number versus sideband Rabi frequency under different experimental parameters.} Each dot is an average phonon number measured in the same way as in Fig.~\ref{Fig 2}, with the error bar representing one standard deviation. In \textbf{a}, \textbf{b} and \textbf{c}, we vary the ratio parameter $R$, the total quench time $\tau_q$ and the motional decoherence time $\tau_d$, respectively, while keeping the other parameters the same as those in Fig.~\ref{Fig 2}. \textbf{a}. We keep $\Omega_{\mathrm{SB}}^\mathrm{c}=2\pi\times10\,\kHz$ and $\tau_q=2\,\ms$. Then we need $\delta_b=2\pi\times41.3\,\kHz$ and $\delta_r=2\pi\times36.1\,\kHz$ for $R=15$ and $\delta_b=2\pi\times26.8\,\kHz$ and $\delta_r=2\pi\times17.9\,\kHz$ for $R=5$. \textbf{b}. We keep $\Omega_{\mathrm{SB}}^\mathrm{c}=2\pi\times10\,\kHz$, $R=25$, but use different quench time $\tau_q$.
   \textbf{c}. We keep $\Omega_{\mathrm{SB}}^\mathrm{c}=2\pi\times10\,\kHz$, $R=25$ and $\tau_{q}=2\,\ms$, but vary the motional decoherence time $\tau_d$ by turning on ($\tau_d=5.5\,\ms$) or off ($\tau_d=0.7\,\ms$) the locking of the experimental sequence to the $50\,\Hz$ reference. The curves in \textbf{a} and \textbf{b} are from numerical simulation without considering the motional decoherence, similar to the $\tau_d=\infty$ curve in \textbf{c}. The other two curves in \textbf{c} include the motional decoherence effect by numerically solving a Lindblad master equation (see Methods for more details). The difference between $\tau_d=\infty$ and $\tau_d=5.5\,\ms$ is very small for the quench time $\tau_q=2\,\ms$, thus justifies our simplification of $\tau_d=\infty$ for \textbf{a}, \textbf{b} and the previous numerical simulations.}
   \label{Fig 3}
\end{figure*}

\textbf{Scaling of the order parameter with respect to various experimental parameters.}
Finally we consider the scaling of the order parameter with respect to different experimental parameters. For this purpose, the average phonon number is the preferred observable because it can vary in a wider range than the spin-up state population. Our results are summarized in Fig.~\ref{Fig 3} where we change the ratio parameter $R$, the total quench time $\tau_q$ and the motional decoherence time of the ion $\tau_d$, while keeping the other parameters the same.
Figure~\ref{Fig 3}a considers different ratios $R=(\delta_b+\delta_r)/(\delta_b-\delta_r)$ by keeping the critical sideband Rabi frequency $\Omega_{\mathrm{SB}}^\mathrm{c}=2\pi\times10\,\kHz$ fixed. Hence we can deduce $\delta_{b(r)}=\Omega_{\mathrm{SB}}^\mathrm{c} (\sqrt{R} \pm 1/\sqrt{R})$ from the ratio parameter $R$.
As expected, the sharpness of the curve and the final average phonon number are positively correlated with the ratio parameter, and approach nonanalytical behavior in the limit $R\to\infty$ (see Supplementary Information for a further discussion about the finite-ratio scaling).
In Fig.~\ref{Fig 3}b, we vary the quench time $\tau_q$ to study its effect on the order parameter. A shorter quench time leads to a larger deviation from the adiabatic evolution, thus the prepared state has larger deviation from the true ground state. Only for long enough quench time can the prepared states have large enough overlap with the real ground states, hence show the clear evidences of the QPT.
In Fig.~\ref{Fig 3}c we study the influence of finite motional decoherence time $\tau_d$ of the trapped ion. To keep the quantum nature of the system during the slow quench dynamics, the quench time should be within the coherence time of the system. As is mentioned above, the motional coherence of our trap is largely affected by the $50\,\Hz$ noise from the AC power line. By locking the experimental sequence to the $50\,\Hz$ reference, the coherence time is above $5\,\ms$; while if we turn off the locking, the coherence time will drop below $1\,\ms$. This phenomenon is also reported in Ref.~\cite{Henning-Kaufmann}. We conduct the experiments with the locking turned on and off, respectively. As expected, the sharpness of the curve reduces for shorter coherence time. The results agree well with the theoretical prediction for a motional decoherence time $\tau_d=5.5\,\ms$ and $\tau_d=0.7\,\ms$ respectively from solving the Lindblad master equation (see Methods). We also perform a simulation using the Schr\"odinger equation without considering any decoherence, which is labelled as $\tau_d=\infty$. This curve is very close to that for $\tau_d=5.5\,\ms$, which justifies our numerical simulation using Schr\"odinger equation in Figs.~\ref{Fig 1}, \ref{Fig 2}a, \ref{Fig 3}a and \ref{Fig 3}b, for $\tau_q\le 2\,\ms$.
We also notice that the heating of the motional mode and the decoherence of the qubit state are potential sources of errors in the experiments. However, in our system the heating rate measured by the standard method \cite{motional-heating-rate,PhysRevA.61.063418} is well below $50\,$quanta/s and the qubit coherence time measured by the Ramsey method using Raman transition is greater than $50\,\ms$ which is mainly limited by the coherence of the PLL \cite{PLL}.
Both have negligible effect on the measured order parameters as discussed in Methods.

\section{Discussion}
To sum up, we have successfully observed a QPT from the normal phase to the phonon superradiance phase associated with the QRM simulated by a single trapped ion. Through slow quench dynamics, we measure the spin-up state population and the average phonon number as the order parameters and observe them changing from near zero to large values when the control parameter is tuned across the quantum critical point. For the average phonon number, the change becomes sharper when the ratio parameter increases, analogous to approaching closer to a thermodynamic limit. The strong controllability of the trapped-ion system also allows us to vary the experimental parameters and study their influence on the phase transition. We also note that in Ref~\cite{PhysRevLett.118.073001}, a method to observe the universal scaling with spin-up state population was proposed. However, considering some technical difficulties, it is not possible for our system to observe the critical phenomena currently (see Supplementary Information for more discussions about this). To further study the finite-ratio scaling, we will either need to reduce the experimental noise and to upgrade the experimental setup to get more accurate results near the critical point for larger frequency ratio $R$; or we may need to develop different scaling methods which use data points farther away from the critical point. Our work is a first step towards the more detailed studies of the QPT in the QRM, including the critical dynamics and the universal scaling \cite{PhysRevLett.115.180404,PhysRevLett.118.073001}. With reservoir engineering \cite{Myatt2000,Kienzler53}, it is also possible to observe the dissipative phase transition in the QRM \cite{PhysRevA.97.013825}. Besides, our method can be directly extended to study the QPT in the many-body version of the QRM, i.e. the Dicke model \cite{PhysRev.93.99,PhysRevLett.108.043003,PhysRevA.85.043821} when we increase the number of the trapped ions.

\section{Methods}
\textbf{AC Stark shift compensation.}
Our $355\,\nm$ pulsed laser has a frequency comb structure with a repetition rate $\omega_{\mathrm{rep}}\approx 2\pi \times 118.695\,\MHz$ and a bandwidth of about $200\,\GHz$. It can be used to bridge the transition between the two qubit levels with a frequency difference around $\omega_q\approx 2\pi\times 12.6\,\GHz$, without the need of large frequency shifts between the two Raman beams \cite{frequencycomb}. In Fig.~\ref{Fig 0}a, suppose AOM1 introduces a frequency shift of $\omega_{\mathrm{AOM1}}$, which is dynamically varied to compensate the fluctuation of the repetition rate $\omega_{\mathrm{rep}}$ \cite{PLL}, and AOM2 leads to a frequency shift $\omega_{\mathrm{AOM2},r(b)}$ for the red (blue) component of the bichromatic laser beams. The closest differential frequencies to the sideband transitions will be $\Delta\omega_{r(b)}=n \times \omega_{\mathrm{rep}}+\omega_{\mathrm{AOM1}}-\omega_{\mathrm{AOM2},r(b)}$ with $n=107$, the span number of the frequency-comb pairs as shown in Fig.~\ref{Fig 0}c.

As we have mentioned in the main text, when tunning the sideband Rabi frequency from zero to a specific value, the AC Stark shift induced by the off-resonant coupling of the undesired frequency-comb pairs will also increase continuously. This is a common shift to $\delta_r$ and $\delta_b$, which changes $\delta_r+\delta_b$ and hence the ratio parameter $R$. For the $355\,\nm$ pulsed laser we use, when the sideband Rabi frequency is set to $2\pi \times 14.2\,\kHz$, the AC Stark shift can reach over $2\pi \times 10\,\kHz$ measured by the standard Ramsey method \cite{PhysRevLett.90.143602}. Such a large shift has non-negligible effect on the order parameters and must be compensated during the slow quench dynamics. Before each round of experiment, we calibrate the AC Stark shift $\Delta_{\mathrm{ac}}$ under the QRM Hamiltonian with different sideband Rabi frequencies $\Omega_{\mathrm{SB}}$ and fit it according to $\Delta_{\mathrm{ac}} = \alpha \Omega^2_{\mathrm{SB}}$ where $\alpha$ is a proportionality constant.
Then when performing the slow quench experiment, we correct the frequency of the blue (red) component in the bichromatic beams as $\omega_{b(r)}(t)=\omega_{b(r)}(0)+\Delta_{\mathrm{ac}}(t)$, to make the detuning $\delta_{b(r)}$ fixed. This can be realized by phase modulating the driving RF signals on AOM2, which can be conveniently implemented by an AWG as shown in Fig.~\ref{Fig 0}a with a pre-determined waveform loading to its memory. The waveform for the pulse is given by $A(t)\cos(\omega_{\mathrm{AOM2},r(b)} t-\int_{0}^t\Delta_{\mathrm{ac}}(t) \mathrm{d}t)$,
where $\omega_{\mathrm{AOM2},r(b)}$ is a pre-set driving frequency of AOM2 at the beginning of the experiment and the driving amplitude $A(t)\propto\Omega_{\mathrm{SB}}(t)$ is also calibrated before the experiment.

\textbf{Phonon number distribution measurement.}
To measure the phonon number of a quantum state of the spin-phonon system, we trace out the spin part by optically pumping it to $\ket{\downarrow}$ \cite{PhysRevA.76.052314} within a duration of $5\,\us$ so that its influence to the motional state can be neglected. Then we apply a blue sideband pulse with various duration $t$ and measure the resultant spin-up state population $P_{\uparrow}(t)$. It can be fitted by \cite{RevModPhys.75.281,PhysRevLett.76.1796,PhysRevX.8.021027}
\begin{equation}
\label{8}
P_{\uparrow}(t)=\frac{1}{2}\left[1 - \sum_{k=0}^{k_{\mathrm{max}}} p_ke^{-\gamma_k t}\cos (\Omega_{k,k+1} t)\right],
\end{equation}
where $p_k$ is the occupation of the phonon number state $\ket{k}$, $\gamma_k$ is a number-state-dependent empirical decay rate of the Rabi oscillation where we adopt a commonly used form $\gamma_k\propto (k+1)^{0.7}$ \cite{RevModPhys.75.281,PhysRevLett.76.1796,PhysRevX.8.021027}, $\Omega_{k,k+1}=\sqrt{k+1}\Omega_{\mathrm{SB}}$ is the number-state-dependent sideband Rabi frequency, and $k_{\mathrm{max}}$ is the cutoff in the phonon number. If the hyperparameter $k_{\mathrm{max}}$ in the fitting model is too small, we will lose the high-phonon population and thus limited to a small average phonon number; however, if $k_{\mathrm{max}}$ is chosen too large, the uncertainty in the fitting will increase because we need to fit more parameters; and the risk of misjudgement of high-phonon population from the noise of the blue-sideband signals will also increase (see Supplementary Information for more details about the choice of $k_{\mathrm{max}}$).

After fitting the phonon state population $P=(p_0,\,p_1,\,\cdots)^T$ with its covariance matrix $\Sigma$, we can compute the average phonon number $\bar{n} = N \cdot P$ where $N=(0,\,1,\,\cdots)$ is a row vector representing the phonon number basis. Assuming the fitted parameters follow a joint Gaussian distribution \cite{Nonlinear} (see Supplementary Information for more details about this assumption), we can estimate the variance of $\bar{n}$ as $\sigma_{\bar{n}}^2 = N \Sigma N^T$.

\textbf{Error analysis and numerical simulation.}
To consider the motional decoherence effect, we numerically solve the master equation with the Lindblad superoperator $L[\hat{O}]\hat{\rho} \equiv \hat{O} \hat{\rho} \hat{O}^\dag - \hat{O}^\dag \hat{O}\hat{\rho} / 2 - \hat{\rho} \hat{O}^\dag \hat{O} / 2$ of dephasing type \cite{PhysRevA.62.053807}:
$\dot{\hat{\rho}}(t)=-i[\hat{H}, \hat{\rho}(t)] + L[\sqrt{2\Gamma_m}\hat{a}^\dag\hat{a}]\hat{\rho}$,
where $\Gamma_{\mathrm{m}}=1/\tau_d$ is the dephasing rate with the decoherence time $\tau_d$. In Fig.~\ref{Fig 3}c with the line-trigger on (off), we set $\tau_d=5.5\,\ms$ ($0.7\,\ms$) which is within the range of our daily measurement (see Supplementary Information for more detials about the motional coherence measurement), to fit the experimental data.

For the motional heating and the qubit decoherence, we add the Lindblad superoperators $L[\sqrt{\gamma n_{\mathrm{th}}}\hat{a}^\dag]+L[\sqrt{\gamma (n_{\mathrm{th}}+1)}\hat{a}]$ \cite{PhysRevA.62.053807} and $L[\sqrt{2\Gamma_q}\hat{\sigma}_{+}\hat{\sigma}_{-}]$ \cite{quantum-optics}, respectively, where $\gamma n_{\mathrm{th}}\approx\gamma (n_{\mathrm{th}}+1)$ is the motional heating rate which is below $50\,\mathrm{s}^{-1}$ and $\Gamma_{q}$ is the qubit decoherence rate which is below $20\,\mathrm{s}^{-1}$ in our system. As we have mentioned in the main text, the effects of these two terms are negligible from numerical simulation. All the Lindblad superoperators we used in the master equation just represent the results in the lab frame (describing the experimental decay), and does not represent decay in the simulated system frame (describing the QRM decay).

The fluctuation of the trap frequency (motional mode frequency $\omega_x$), which is within $2\pi\times\,150\Hz$ after applying the RF power stabilization \cite{doi:10.1063/1.4948734}, can be the main error source on the ratio parameter $R$, because the trap frequency fluctuation is asymmetrical for $\delta_r$ and $\delta_b$ (see Fig.~\ref{Fig 0}d), causing $\delta_b-\delta_r$ to change, thus the ratio parameter. Under $2\pi\times150\,\Hz$ trap frequency fluctuation, the uncertainty for $R=25,\,15,\,5$ are $\pm1.7$, $\pm0.82$ and $\pm0.16$, respectively.
Other sources of errors can be from the phonon number fitting beacuse some noise in the blue-sideband signals may be incorrectly recognized as a high-phonon population and cause the fitting error; and from the fluctuation of the AC Stark shift due to the fluctuation of the laser repetition rate and the laser intensity. Consider a $1\%$ sideband Rabi freqeuncy fluctuation (i.e. $1\%$ of $2\pi\times14.2\,\kHz$ for maximal estimation) and $2\pi\times30\,\Hz$ fluctuation of the repetition rate, the standard deviation of the fluctuated AC Stark shift from a theoretical calculation \cite{PhysRevA.94.042308} can reach about $2\pi\times400\,\Hz$. Under this value, the ratio parameter uncertainty for $R=25,\,15,\,5$ are $\pm0.20$, $\pm0.15$ and $\pm0.09$, respectively.

\bigskip

\textbf{Data Availability:} The data that support the findings of this study are available from the corresponding authors upon reasonable request.

\textbf{Code Availability:} The code used for numerical simulations is available from the corresponding authors upon reasonable request.

\textbf{Acknowledgements:}  This work was supported by the National key Research and Development Program of China (2016YFA0301902), the Beijing Academy of Quantum Information Sciences, the Frontier Science Center for Quantum Information of the Ministry of Education of China, and the Tsinghua University Initiative Scientific Research Program. X.Z. acknowledges in addition
support from the National Natural Science Foundation of
China (11704408, 91836106) and the Beijing Natural Science Foundation (Z180013). Y.K.W. acknowledges support from Shuimu Tsinghua Scholar Program and the International Postdoctoral Exchange Fellowship Program.

\textbf{Competing interests:} The authors declare that there are no competing interests.

\textbf{Author Information:} Correspondence and requests for materials should be addressed to L.M.D.
(lmduan@tsinghua.edu.cn).

\textbf{Author Contributions:} L.M.D. proposed and supervised the project. M.L.C., W.D.Z., Q.X.M., Y.J., L.H., X.Z., Z.C.Z. carried out the experiment. Z.D.L. and Y.K.W. carried out the theoretical analysis. M.L.C., Y.K.W., and L.M.D. wrote the manuscript.

\section{Supplementary Information}

\subsection{Note on the error bar estimation} 
\textbf{Error bar estimation in the spin population experiment.}
In the spin population experiment, there are mainly two types of experimental noise we are considering: one is the intrinsic quantum fluctuation and the other is the extrinsic fluctuation of control parameters and environmental parameters. During one round of the experiment, the system is relatively stable and we are mainly concerned with the quantum projection noise \cite{PhysRevA.47.3554}. It arises because the quantum state is not an eigenstate of the observable, say, the spin-up state population and thus by repeating the experiment we get different outcomes even if we prepare the same quantum state. This noise can be suppressed by increasing the number of measurements. By averaging over 500 shots in each experimental round, we get the average spin-up state population with the quantum projection noise suppressed to $1/\sqrt{500}$, which is small compared with other experimental noise.

On the other hand, the prepared quantum states can differ due to the long-term fluctuation of control parameters and environmental parameters. This noise cannot be suppressed by increasing the number of measurements and we regard this as the dominant error source in our experiment. These effects include fluctuation in laser intensity, laser repetition rate, temperature, air pressure, etc. Therefore, we conduct the experiment for 20 rounds, each at a different time with the time interval on the order of several minutes. We then use the standard deviation of the 20-round outcomes to estimate the error bar.

\textbf{Error bar estimation in the phonon number experiment.}
When estimating the error bar of the average phonon number, we need to make an assumption about the distribution of the experimental noise. Under the common assumption of independent and identically distributed Gaussian noise of the experimental data, it can be shown that the fitted parameters also follow a joint Gaussian distribution (see e.g. Theorem 2.1 of Ref.~\cite{Nonlinear}.), which is what we use in this work. We want to emphasize that this assumption is used in lots of experiments when extracting parameters by fitting the experimental data, and is implicitly used in many scientific computing softwares like MATLAB when fitting parameters.

\subsection{Note on the choice of the $k_{\mathrm{max}}$ in the phonon number distribution fitting}
\begin{figure*}[tbp]
  \centering
  \includegraphics[width = 0.8\linewidth]{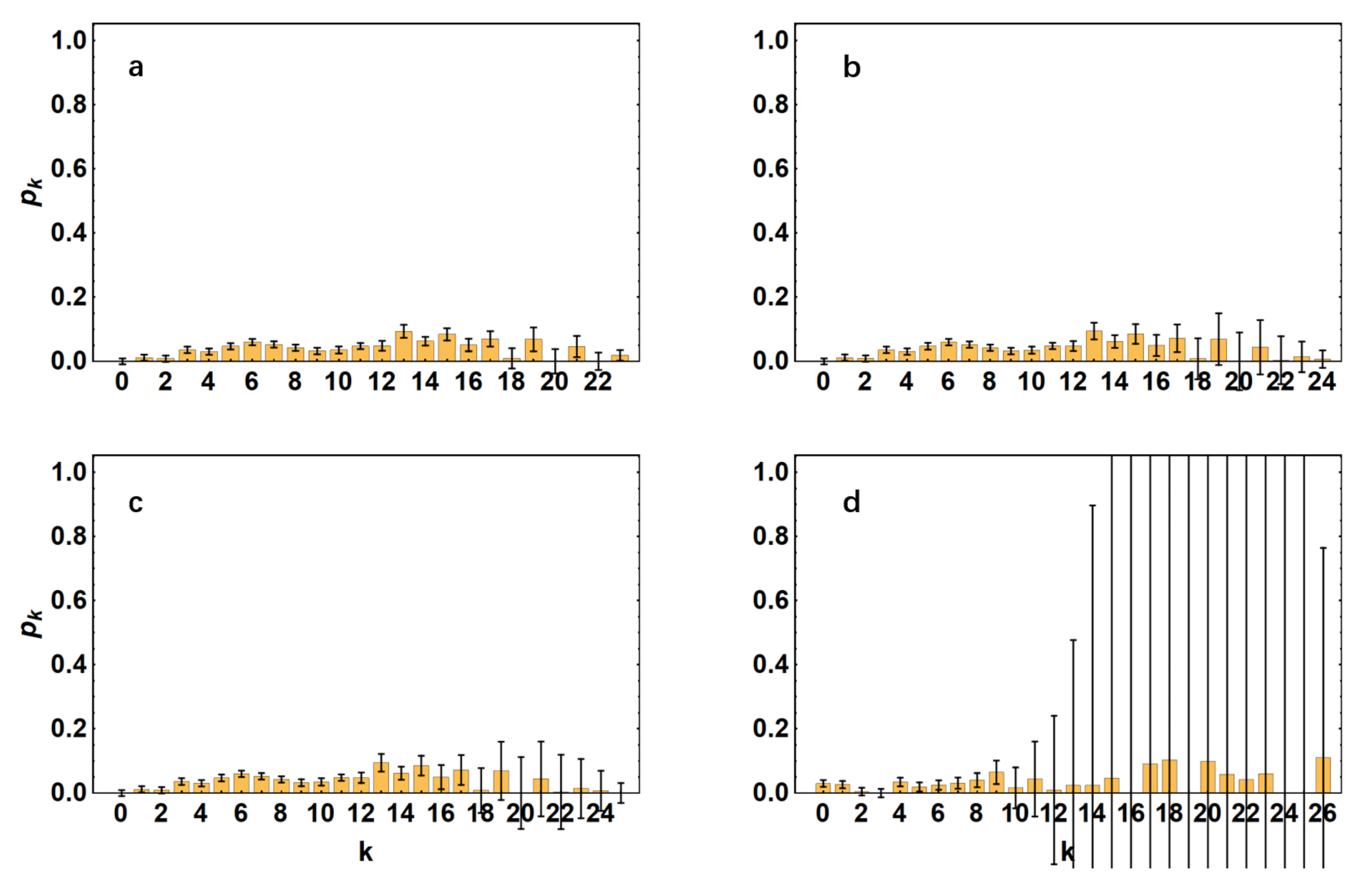}
  \caption{\textbf{Phonon number distribution with different cutoff number $k_{\mathrm{max}}$.} The phonon number distribution with cutoff number $k_{\mathrm{max}}=23$ in \textbf{a}, $k_{\mathrm{max}}=24$ in \textbf{b}, $k_{\mathrm{max}}=24$ in \textbf{c} and $k_{\mathrm{max}}=24$ in \textbf{d}. The error bar is one standard deviation from the fitting program. The extracted average phonon number is $11.54\pm0.71$ while the total occupation $\sum_{k=0}^{k_{\mathrm{max}}} p_k$ is around $95.6\%$ in \textbf{a}. The results of the phonon number distribution are nearly the same with the total occupation around $95.9\%$ and the average phonon number $11.63\pm0.73\,(0.74)$ in \textbf{b} and \textbf{c}. However, when the cutoff number is set to 26 (\textbf{d}), the phonon number distribution dramatically changes and the error bar of the occupation of the Fock states after $\ket{12}$ becomes very large, indicating that overfitting occurs.}
  \label{Fig:phonon number distribution}
\end{figure*}

We use the lowest cutoff number that can ensure the total occupation of all the Fock states to be above $95\%$ as $k_{\mathrm{max}}$ in the phonon number distribution fitting. We take the phonon number distribution of the state with the largest average phonon number in this experiment as an example to show how we choose a proper $k_{\mathrm{max}}$. In Fig.~\ref{Fig:phonon number distribution}a, the extracted average phonon number is $11.54\pm0.71$ while the total occupation $\sum_{k=0}^{k_{\mathrm{max}}} p_k$ is around $95.6\%$ with a cutoff number 23 (which can be seen from the horizontal axis). When we continue to increase the cutoff number to 24 (Fig.~\ref{Fig:phonon number distribution}b) and 25 (Fig.~\ref{Fig:phonon number distribution}c), the results of the phonon number distribution are nearly the same with the total occupation around $95.9\%$ and the average phonon number $11.63\pm0.73\,(0.74)$. However, when the cutoff number is set to 26 (Fig.~\ref{Fig:phonon number distribution}d), the phonon number distribution dramatically changes and the error bar of the occupation of the Fock states after $\ket{12}$ becomes very large, indicating that overfitting occurs. Also, according to the numerical simulation, the total occupation number above the Fock state $\ket{24}$ (including $\ket{24}$) is only $0.13\%$, contributing an average phonon number around 0.03 to this state, which is much smaller than the error due to the fitting of about 0.7. Hence, this also justifies the choice of the cutoff number 23.
The reason why the total occupation in our fitting is only around $95\%$ may be due to the state preparation and measurement error (SPAM error) during the blue-sideband pulse analysis (described in Methods). For instance, even if we can ideally prepare the phonon ground state, i.e. only the Fock state $\ket{0}$ is occupied and the only non-zero occupation is $p_0$. We can easily see that $p_0$ is the contrast of the sinusoidal spin-up state population curve used to extract the occupation number. However, due to the SPAM error, the contrast must be less than 1. In our system, the SPAM error is around $2\%$ (an average of $1\%$ dark-state detection error and $3\%$ bright-state detection error), which means the contrast of the spin-up state population curve is only $96\%$. This explains the relatively low total occupation.

\subsection{Note on the correction for the Lamb-Dicke approximation}
All of our discussions in the main text are based on the condition that the single trapped ion is in the Lamb-Dicke regime. In this regime, the extension of the ion's wave function is much smaller than the laser's wavelength, or this limitation can be written as $\eta\sqrt{2\bar{n}+1} \ll 1$ \cite{RevModPhys.75.281}, where $\eta$ is the Lamb-Dicke parameter and $\bar{n}$ is the average phonon number of the motional state. In our system, the Lamb-Dicke parameter is around $0.07$. However, in our experiment, the maximum average phonon number exceeds ten, which means $\eta\sqrt{2\bar{n}+1}$ is around $0.3$, making the non-linear terms of $\eta$ a non-negligible effect to the entire model Hamiltonian. In the following, we consider the corrections to the numerical results of the two order parameters due to the non-linear effect.

When we consider the non-linear terms, the total Hamiltonian of the QRM simulated by a single trapped ion reads \cite{PhysRevA.97.023624}:
\begin{equation}
    \hat{H}_{\mathrm{NQRM}}=\frac{\omega_{\mathrm{a}}}{2} \hat{\sigma}_{z}+\omega_{\mathrm{f}} \hat{a}^{\dagger} \hat{a}+\lambda\left(\hat{\sigma}_{+}+\hat{\sigma}_{-}\right)\left(\hat{f} \hat{a}+\hat{a}^{\dagger} \hat{f}\right),
\end{equation}
where the non-linear effect is embodied in the function \cite{PhysRevA.52.4214} 
\begin{equation}
    \hat{f}(\hat{a},\hat{a}^\dag)=e^{-\eta^{2} / 2} \sum_{l=0}^{\infty} \frac{\left(-\eta^{2}\right)^{l}}{l !(l+1) !} \hat{a}^{\dagger l} \hat{a}^{l}.
\end{equation}
When we only consider the first expansion term, i.e. $l=0$ and neglect the term $e^{-\eta^{2}/2}$, the Hamiltonian reduces to the linear QRM. Here, we implement a numerical simulation additionally considering an $l=1$ term.    

\begin{figure*}[tbp]
  \centering
  \includegraphics[width = 0.95\linewidth]{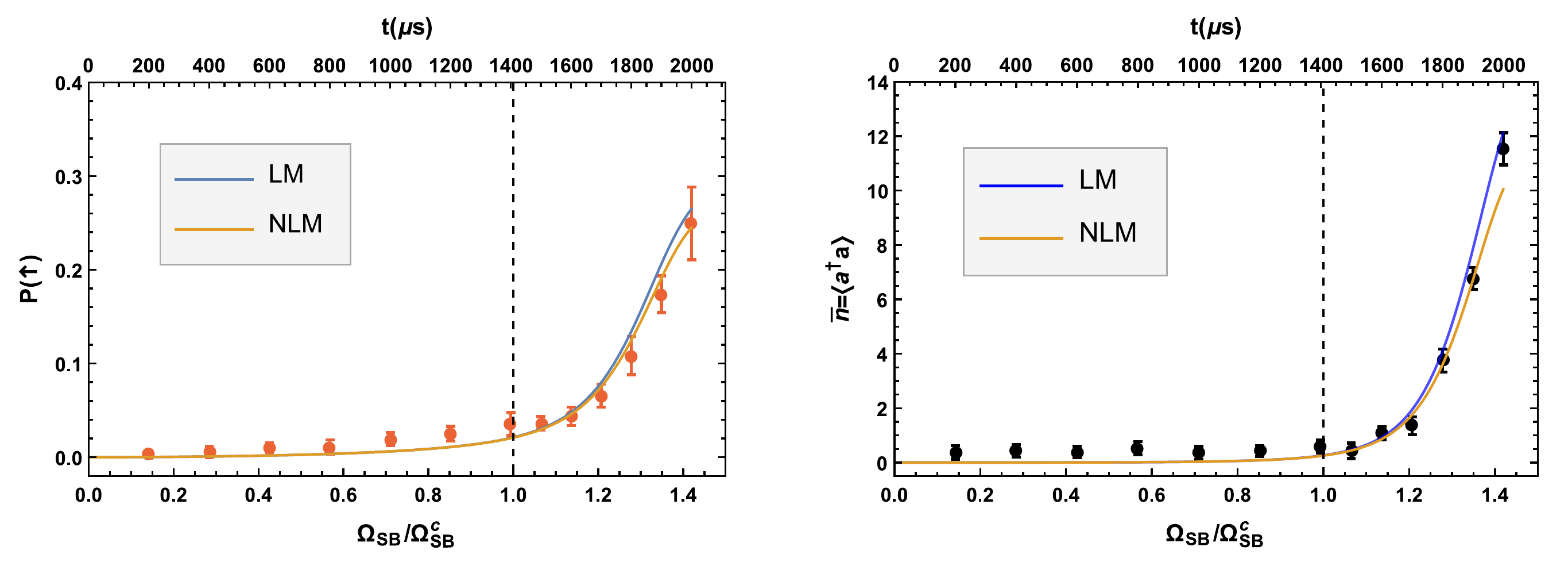}
  \caption{\textbf{The spin-up state population and the average phonon number versus the sideband Rabi frequency with/without non-linear effect.} Here we set the experimental parameters the same as the main text with $\delta_{b}=2 \pi \times 52.0\, \mathrm{kHz}$, $\delta_{r}=2 \pi \times 48.0\, \mathrm{kHz}$, thus the ratio parameter $R=25$. The total quench time $\tau_q=2\,\mathrm{ms}$ with the sideband Rabi frequency increases linearly from zero to $\Omega_{\max }=2 \pi \times 14.2\, \mathrm{kHz}$. The Lamb-Dicke parameter is $\eta=0.07$. \textbf{a} and \textbf{b} are Fig.~2 and Fig.~3a in the main text with an additional numerical result of the non-linear QRM, respectively. We can see clearly that in the normal phase, the phonon number is small enough that both the two order parameters in the non-linear model (NLM) show good consistency with those in the linear model (LM). In the superradiant phase, with the increase of the average phonon number, the non-linear effect becomes more and more significant and causes a non-negligible deviation of the two order parameters between the NLM and the LM.}
  \label{fig:nonlinear_QRM} 
\end{figure*}

As shown in Fig.~\ref{fig:nonlinear_QRM}, with the same experimental parameters as in the main text, we simulate the effect on the spin-up state population and the average phonon number during the quench dynamics. As we can see, in the normal phase, the phonon number is small enough that both the two order parameters in the non-linear model (NLM) show good consistency with those in the linear model (LM). In the superradiant phase, with the increase of the average phonon number, the non-linear effect becomes more and more significant. In our simulation, we find that the maximum relative deviation of the average phonon number between the NLM and the LM ($|\bar{n}_{\mathrm{LM}}-\bar{n}_{\mathrm{NLM}}|/\bar{n}_{\mathrm{LM}}$) is about $17\,\%$. However, the deviation near the critical point is only about $2\,\%$, which is small enough compared with other errors discussed in the Methods. 

In conclusion, the non-linear terms in the simulated QRM causes a small but non-negligible deviation when the average phonon number is large ($\gtrsim 10$). However, because they are still smaller than the leading term, we expect the qualitative behavior of the quantum phase transition, in particular the universal class near the phase transition point, to be unaffected.

\subsection{Note on the scaling analysis}
\textbf{Scaling analysis with spin population.}
We note in Ref.~\cite{PhysRevLett.118.073001}, spin population is used to analyze the scaling effect of the QPT in the QRM. However, some of the experimental parameters and conditions in Ref.~\cite{PhysRevLett.118.073001} are rather stringent for our system. There are mainly three conditions that are currently not achievable in our system. First, in Ref.~\cite{PhysRevLett.118.073001} the bosonic mode frequency $\tilde{\omega}_{0}/2\pi$ ($\omega_\mathrm{f}/2\pi$ in our notation) is set to $200\,\Hz$ to realize large frequency ratio $R$ of 50 to 400 under realistic coupling strength. This is comparable to the trap frequency fluctuation (around $150\,\Hz$) and even smaller than the fluctuation of the estimated AC Stark shift (around $400\,\Hz$, see Methods) in our system, and therefore will lead to large error. Second, under such large frequency ratios, the required adiabatic evolution time of about $250\,\ms$ is too long compared to our qubit coherence time under Raman laser of about $40$ to $60\,\ms$. Finally, Ref.~\cite{PhysRevLett.118.073001} proposes a standing wave configuration for the laser beams in order to suppress the influence of the carrier term under large frequency ratio, but our setup uses a traveling wave configuration which is more common in current ion trap experiments. The standing wave configuration needs four laser beams instead of the two beams in the traveling wave configuration. It is not easy to change our current configuration to four beams. We believe these technical challenges can be overcome with (1) choosing a more appropriate bosonic mode frequency (e.g $1\,\kHz$) with a still achievable coupling strength ($300\,\kHz$ carrier Rabi rate); (2) suppressing the system noises by improving the RF amplitude stabilization system and choosing a more appropriate repetition rate of the Raman laser; (3) improving the coherence time of the system (including the motional coherence time). 
\begin{figure*}[tbp]
  \centering
  \includegraphics[width = 0.95\linewidth]{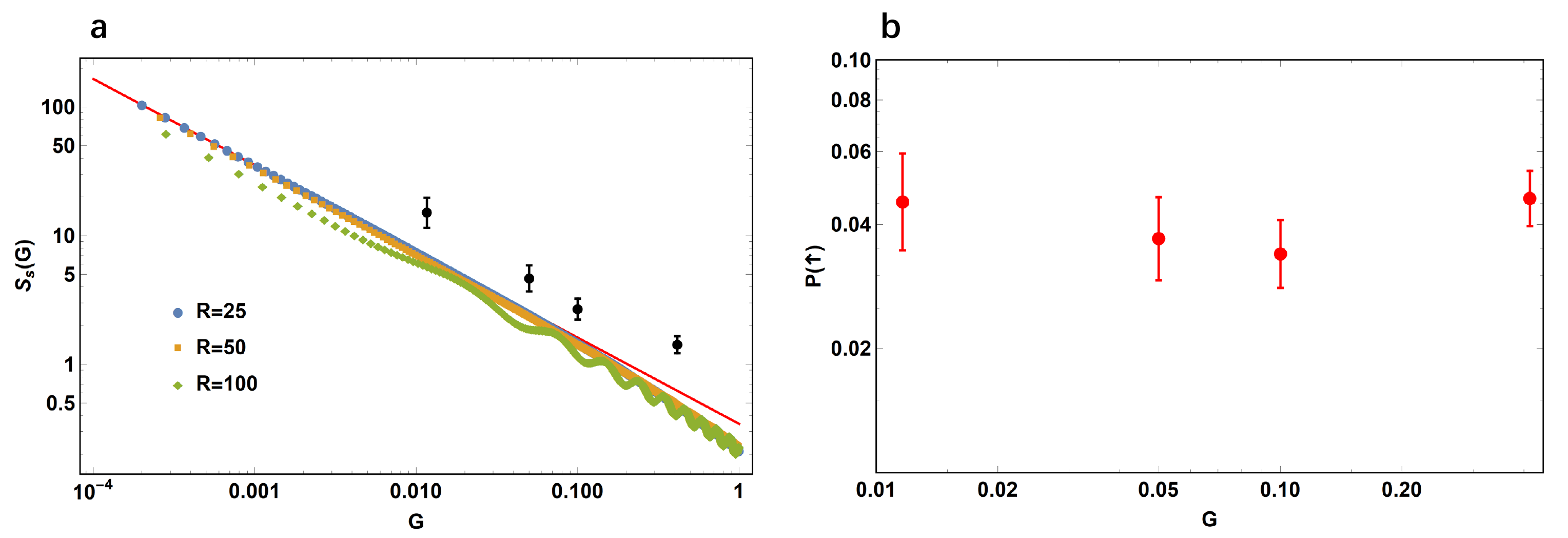}
  \caption{\textbf{Scaling analysis with spin-up state population.} \textbf{a}. The $S_s(G)$-$G$ plot, where $S_s(G)\equiv2P(\uparrow)|g-1|^{-1}$ and $G\equiv R|g-1|^{3/2}$ with $g$ the coupling strength and $P(\uparrow)$ the spin-up state population. The blue, yellow and green points are numerical simulation results and the red curve is an analytical line with a slope $-2/3$, which is a critical exponent \cite{PhysRevLett.118.073001}. The numerical results agree well with the analytical line except the numerical result with $R=100$. This is because when the ratio $R$ is too large, the carrier term in the trapped-ion simulation will cause the simulated Hamiltonian to deviate from the real QRM model \cite{PhysRevLett.118.073001}. The black points with error bar are calculated from the experimental results with $R=25$. The error bar is estimated as the error bar of the spin-up state population $P(\uparrow)$ multiplied by the corresponding $|g-1|^{-1}$. \textbf{b}. The four experimental data presented in \textbf{a} near the critical point $g_\mathrm{c}=1$, with their raw values of $(g,P(\uparrow))$ being $(0.994,0.0453\pm0.0123)$, $(0.984,0.0369\pm0.0085)$, $(0.975,0.0339\pm0.0064)$, $(1.065,0.0462\pm0.0071)$ respectively. Although in the log-plot, these points seem to nicely follow a trend with the red line and their error bars are not that large compared to the difference of these points, the difference between the raw data points is on the same order as the raw data error bars. We believe the trend indicated by the four black points is just dominated by the dependence of $S_s(G)$ and $G$ on $|g-1|$.}
  \label{Fig:spin scaling}
\end{figure*}

Given the current condition of our system, we choose a moderate ratio $R=25$ to implement the spin population experiment and show the overall behavior in Fig.~2 in the main text. Here we further supplement some experimental data around the critical point together with a numerical simulation according to Ref.~\cite{PhysRevLett.118.073001}. We summarize the results in Fig.~\ref{Fig:spin scaling}a. The figure is a $S_s(G)$-$G$ plot where $S_s(G)\equiv 2P(\uparrow)|g-1|^{-1}$ and $G\equiv R|g-1|^{3/2}$ with $g$ the coupling strength and $P(\uparrow)$ the spin-up state population. The blue, yellow and green points are numerical simulation results and the red curve is an analytical line with a slope $-2/3$ (note that according to Ref.~\cite{PhysRevLett.118.073001}, the asymptotic behavior of $S_s(G)$ is $\lim_{G \to 0}S_s(G) \propto G^{-2/3}$, i.e. there is a universal critical exponent -2/3). The numerical results agree well with the analytical line except the numerical result with $R=100$. This is because when the ratio $R$ is too large, the carrier term in the trapped-ion simulation will cause the simulated Hamiltonian to deviate from the real QRM model and this is why Ref.~\cite{PhysRevLett.118.073001} propose a standing-wave laser configuration to suppress the influence of the carrier term. The black points with error bar are calculated from the experimental results with $R=25$. The error bar is estimated as the error bar of the spin-up state population $P(\uparrow)$ (which is the raw data taken from the experiment) multiplied by the corresponding $|g-1|^{-1}$ which is supposed to be accurate. The four experimental points (from left to right) are all very close to the critical point $g_\mathrm{c}=1$, where their raw data values of $(g,P(\uparrow))$ are $(0.994,0.0453\pm0.0123)$, $(0.984,0.0369\pm0.0085)$, $(0.975,0.0339\pm0.0064)$, $(1.065,0.0462\pm0.0071)$ respectively. Although in the log-plot, these points seem to nicely follow a trend with the red line and their error bars are not that large compared to the difference of these points, the difference between the raw data points is on the same order as the raw data error bars (see Fig.~\ref{Fig:spin scaling}b). We believe the difference between the raw data points can be easily washed out due to experimental noises (e.g. the fluctuations of trap frequency and AC Stark shift) because they are too close to the same point. Hence, we believe the trend indicated by the four black points is just dominated by the dependence of $S_s(G)$ and $G$ on $|g-1|$. In conclusion, the precision of the current experiment prevents us from observing the universal scaling law with spin population.

\begin{figure}[tbp]
  \centering
  \includegraphics[width=0.95\linewidth]{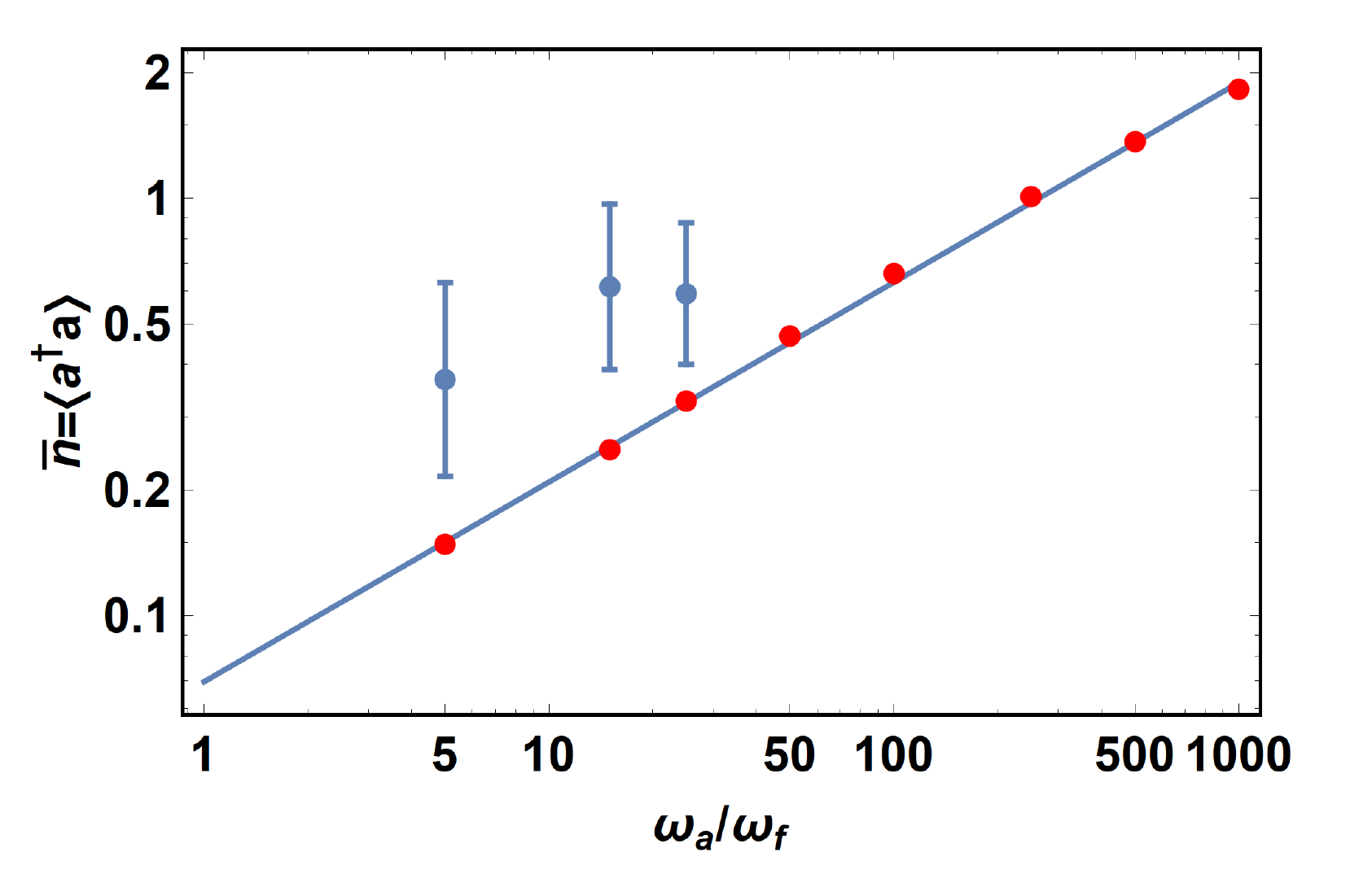}
  \caption{\textbf{Finite-ratio scaling of the average phonon number near the critical point.} The red points are the numerical results with the system size indicated by the ratio $R=\omega_\mathrm{a}/\omega_\mathrm{f}$ ranging from 5 to 1000 and the fitting result shows that the slope of the fitting linear line is 0.48. The blue points with error bar are the experiment results. Under the current achievable ratio $R$, the difference between these points is on the same order of magnitude as the error bar, indicating they are vulnerable to the experimental noises.}
  \label{fig:phonon scaling}
\end{figure}

\textbf{Scaling analysis with average phonon number.}
We present a numerical simulation of the finite-ratio scaling of the average phonon number near the critical point $g_\mathrm{c}=1$ and show the result in Fig.~\ref{fig:phonon scaling}.
The red points are the numerical results with the system size (indicated by the ratio $R\equiv\omega_\mathrm{a}/\omega_\mathrm{f}$) ranging from 5 to 1000 and the fitting result shows that the slope of the fitting line is 0.48. The blue points with error bar are the experimental results. Under the current achievable ratio $R$, the difference between these points is on the same order of magnitude as the error bar, indicating they are vulnerable to the experimental noises. Thus these points cannot be used to extract the critical exponent. Also, we note that the fitted slope of 0.48 from the numerical simulation data actually deviates from the true critical exponent 1/3 in the regime $R\to \infty$ in analytic calculation (see Ref.~\cite{PhysRevLett.115.180404}). In order to see this precise exponent, the ratio $R$ in the numerical simulation needs to exceed $10^5$. Due to such large ratio, the adiabatic ground state preparation may need a duration orders of magnitude larger than the coherence time of the system. Hence it is not achievable for our system currently to observe the precise scaling effect and to extract the critical exponent with average phonon number. We can only observe the overall behavior of the phonon number variation curves with three different ratios (5, 15 and 25), and as expected the curve becomes sharper with larger ratio (see Fig.~4a in the main text).

\subsection{Note on the Ramsey interferometric measurement for motional coherence} 
\begin{figure*}[tbp]
  \centering
  \includegraphics[width = 0.95\linewidth]{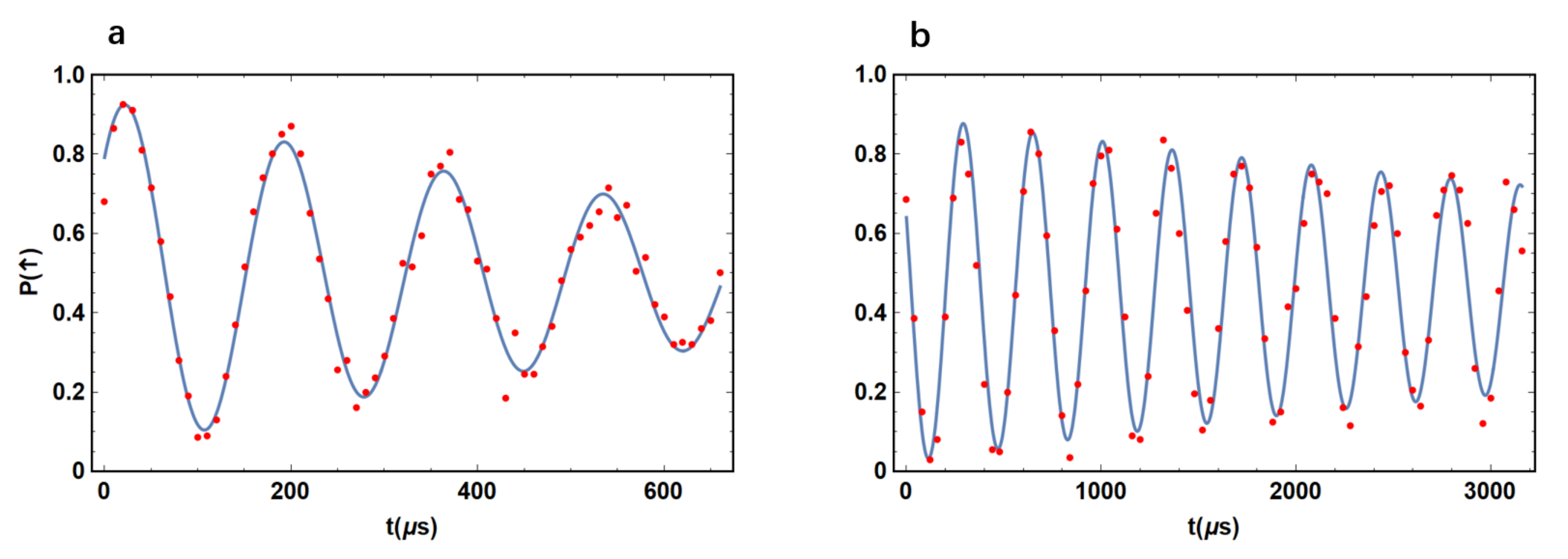}
  \caption{\textbf{The motional coherence time measured by the Ramsey method.} \textbf{a}. Without the line-trigger on, the Ramsey fringes decay fast and the estimated coherence time is around 0.7 ms. \textbf{b}. With the line-trigger on, the Ramsey fringes decay much slower and the estimated coherence time is around 5.5 ms}
  \label{fig:motional coherence} 
\end{figure*}

We use the commonly used Ramsey method \cite{PhysRev.78.695} to measure the motional coherence time with or without the line-trigger on. We apply two $pi/2$ blue-sideband pulses with a time interval $\tau$ in between and then measure the spin population. By varying the time interval $\tau$, we obtain the Ramsey fringes shown in Fig.~\ref{fig:motional coherence}. We fit the result by an attenuated sinusoid curve $A e^{-t/\tau_d} \cos (\omega t + \phi)$ where $A$, $\tau_d$, $\omega$ and $\phi$ are the fitting parameters. The coherence time $\tau_d$ is extracted from the fitted curve. In Fig.~\ref{fig:motional coherence}a, the estimated coherence time is around $0.7\,\ms$ and in b, the estimated coherence time is around $5.5\,\ms$. As we can see, the line-trigger can significantly improve the motional coherence time.

\end{document}